%% file: arxiv.tex
\documentclass{article}

\usepackage{arxiv}

\usepackage[utf8]{inputenc} 
\usepackage[T1]{fontenc}    
\usepackage[hidelinks,pdfencoding=auto]{hyperref}       
\usepackage{url}            
\usepackage{booktabs}       
\usepackage{amsfonts}       
\usepackage{amsmath}
\usepackage{amssymb}
\usepackage{nicefrac}       
\usepackage{microtype}      
\usepackage{lipsum}		
\usepackage{graphicx}
\usepackage{natbib}
\usepackage{doi}

\usepackage{setspace}
\usepackage{todonotes}

\usepackage[noabbrev,capitalise]{cleveref}
\usepackage{mathtools}

\usepackage{algpseudocode}
\usepackage{algorithm}

\DeclareMathOperator*{\argmin}{arg\,min}
\DeclareMathOperator*{\argmax}{arg\,max}

\newcommand{\maybetexorpdf}[2]{\texorpdfstring{#1}{#2}}

\newcommand{\captionheading}[1]{\textit{#1}}
\newcommand{\figwidth}{6in}
\newcommand{\ignoreme}[1]{}

\newcommand{\theestimator}{{\hat \theta^{(s)}}}
\newcommand{\strawestimator}{{\hat \theta^{(v)}}}
\newcommand{\anembedding}{\epsilon}

\newcommand{\optionalperiod}{.}

\title{\input{thetitle.tex}}
\date{}

\author{
\addorc{}, 
\addorc{}}

\author{ \href{https://orcid.org/0000-0002-4751-8910}{\includegraphics[scale=0.06]{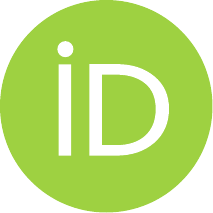}\hspace{1mm}Jackson Loper} \\
	Department of Statistics\\
	University of Michigan\\
	\texttt{jaloper@umich.edu} \\
        \And
        \href{https://orcid.org/0000-0002-1802-4145}{\includegraphics[scale=0.06]{orcid.pdf}\hspace{1mm}Noam Solomon}\\
        Immunai Inc.\\
	\And
	\href{https://orcid.org/0000-0002-1472-5235}{\includegraphics[scale=0.06]{orcid.pdf}\hspace{1mm}Jeffrey Regier} \\
	Department of Statistics\\
	University of Michigan\\
}

\renewcommand{\headeright}{}
\renewcommand{\undertitle}{}
\renewcommand{\shorttitle}{Improving Accuracy in Cell-Perturbation Experiments}

\hypersetup{
pdftitle={A template for the arxiv style},
pdfsubject={q-bio.NC, q-bio.QM},
pdfauthor={David S.~Hippocampus, Elias D.~Striatum},
pdfkeywords={First keyword, Second keyword, More},
}

\begin{document}
\maketitle

\begin{abstract}
{\input{abstract}}
\end{abstract}


\input{main}

\bibliographystyle{plainnat}
\bibliography{refs}

\end{document}

%% file: thetitle.tex
Improving Accuracy in Cell-Perturbation Experiments by Leveraging Auxiliary Information

%% file: abstract.tex
Modern cell-perturbation experiments expose cells to panels of hundreds of stimuli, such as cytokines or CRISPR guides that perform gene knockouts. These experiments are designed to investigate whether a particular gene is upregulated or downregulated by exposure to each treatment. However, due to high levels of experimental noise, typical estimators of whether a gene is up- or down-regulated make many errors. In this paper, we make two contributions.  Our first contribution is a new estimator of regulatory effect that makes use of Gaussian processes and factor analysis to leverage auxiliary information about similarities among treatments, such as the chemical similarity among the drugs used to perturb cells.  The new estimator typically has lower variance than unregularized estimators, which do not use auxiliary information, but higher bias.   To assess whether this new estimator improves accuracy (i.e., achieves a favorable trade-off between bias and variance), we cannot simply compute its error on heldout data as ``ground truth'' about the effects of treatments is unavailable.  Our second contribution is a novel data-splitting method to evaluate error rates.  This data-splitting method produces valid error bounds using ``sign-valid'' estimators, which by definition have the correct sign more often than not.  Using this data-splitting method, through a series of case studies we find that our new estimator, which leverages auxiliary information, can yield a three-fold reduction in type S error rate.

%% file: main.tex
\section{Introduction}

High-throughput cell-perturbation experiments are a transformative way to study cellular biology \citep{schmidt2022crispr}.  Each experimental trial investigates the effect of hundreds of thousands of distinct treatments (also called ``perturbations'') on the expression or thousands of genes in multiple distinct populations of cells (e.g., T cells).  Each trial measures gene expression in treated (perturbed) and control (unperturbed) cells.   The average treatment effect (ATE) for each treatment-gene pair can then be estimated simply by calculating the difference in the gene's expression between cells that received the treatment and the control cells.  However, ATE estimates formed this way often have the incorrect sign: there are treatment-gene pairs for which those estimates indicate that the treatment causes upregulation whereas in fact the treatment caused downregulation, and vice-versa \citep{qiu2020bayesian}.  Sign errors such as these are known as type S errors \citep{gelman2000type}. 
 Type S errors are particularly problematic for cell-perturbation experiments due to the manner in which the ATEs are interpreted.  Specifically, changes in gene expression are often understood in terms of known gene control circuits.  These circuits, in turn, are understood in terms of whether they upregulate and downregulate genes \citep{davidson2001genomic}.  Mistaking upregulation for downregulation, or vice-versa, therefore leads to an incorrect interpretation of the gene control circuits involved in a change.  
 
In this paper, we make two distinct yet complementary contributions.  Our first contribution is a new estimator that reduces type S error by incorporating auxiliary information about similarities between treatments (Section~\ref{sec:estimator}).  For consistency with the literature on spatial statistics, we refer to it as a ``smoothed'' estimator.   We define the smoothed estimator using Gaussian processes, in part because Gaussian processes have already proven effective at incorporating auxiliary information about chemical similarities in other contexts.  For example, \citet{bauer2019gaussian} used Gaussian processes to regress the hydrogen bond acceptor strength of small molecules against their chemical structure.  We further show how low-rank assumptions, inspired by factor analysis, can be incorporated into the Gaussian process model, leading to specialized kernels that increase estimator accuracy.  

Smoothing can reduce error by reducing estimator variance.  However, reducing variance does not necessarily reduce error, as a smoothed estimator may lie in an unfavorable position in the bias-variance trade-off.  Ideally, we would be able to assess estimator error using known parameters of interest that are representative of other parameters of interest, e.g., knowledge of the ATE of some treatments on some genes.  We could then assess whether smoothing increases accuracy.  However, ground truth about ATEs is unavailable in this context.  An alternative method to assess sign error is needed.

Our second contribution is precisely that: a data-splitting method to assess sign error rates without access to ``ground truth'' ATEs (Section~\ref{sec:datasplit}).  This method only requires access to an estimator with a ``sign-validity'' property: the estimator's median must have the same sign as the estimand.  The sign-valid estimators we use in applications are unregularized and have high error rates compared with our smoothed estimator.  We evaluate the smoothed estimates on several experimental replicates and the sign-valid estimator on other experimental replicates.  By comparing the resulting smoothed estimates with the sign-valid estimates, we produce bounds on the number of sign errors made by the smoothed estimator.   

Critically, this procedure bounds the type S errors of an estimator even if that estimator is based on a misspecified model---as our smoothed estimator certainly is, to some extent.  Indeed, our smoothed estimator is constructed using a Gaussian model, but the Gaussian assumption may not hold.  For example, in two of our case studies, we assume Mann-Whitney statistics are Gaussian.  The statistics are based on large populations of cells, so this assumption would be appropriate \emph{if} the measurements for each cell were independent.  However, it is well known that slight variations in lab conditions (such as humidity) introduce significant dependencies \citep{stein2015removing}.  More generally, one could use smoothed estimators based on statistics from more sophisticated differential expression models such as edgeR \citep{robinson2010edger}, even if the Gaussianity of such statistics were not guaranteed.  Our data-splitting method can determine whether smoothing leads to lower type S errors regardless of whether a Gaussian assumption is reasonable.

Our data-splitting method proves that smoothing results in low type S error in several real cell-perturbation experiments (Section~\ref{sec:realdata}).  Through simulation studies, we also explore cases where poor hyperparameter choices instead cause our estimator to have higher error rates than a typical unsmoothed estimator (Section~\ref{sec:sims}).  We find that deleterious smoothing can occur if only a single replicate is used to fit hyperparameters. In contrast, multiple replicates provide multiple observations of each estimand, allowing better hyperparameter estimation and leading to lower error rates.   We conclude by considering smoothed estimators that integrate measurements from multiple experimental platforms (\cref{discussion}).  

\ignoreme{
                _   _               _ 
 _ __ ___   ___| |_| |__   ___   __| |
| '_ ` _ \ / _ \ __| '_ \ / _ \ / _` |
| | | | | |  __/ |_| | | | (_) | (_| |
|_| |_| |_|\___|\__|_| |_|\___/ \__,_|

                                }

\section{Smoothed Estimator}

\label{sec:estimator}

Let $\theta^* \in \mathbb{R}^{P \times G}$ denote a matrix of parameters of interest.  Each entry in this matrix, $\theta_{p,g}^*$, indicates the average treatment effect (ATE) of treatment $p$ on gene $g$. Suppose an experimental procedure to estimate $\theta^*$ has been performed $R$ times, yielding $R$ noisy measurements of each parameter. For $r = 1,\ldots,R$, let $X^{(r)}_{p,g}$ denote the measurement of parameter $(p,g)$ from replicate $r$.  In typical cell-perturbation experiments, $R$ is two or three \citep{subramanian2017next,srivatsan2020massively,schmidt2022crispr}. 

To construct a smoothed estimator, we posit a latent random variable $\theta\in \mathbb{R}^{P \times G}$ and model its distribution using a Gaussian distribution.  We also model the distribution of $X$ given $\theta$ using a Gaussian distribution.  We then produce a smoothed estimator for $\theta^*$ using the posterior expectation, i.e., $\theestimator = \mathbb{E}[\theta |X]$.  

To define the prior distribution of $\theta$, we use three hyperparameters: $K \in \mathbb{R}^{P \times P}$ denotes a matrix that encodes similarities among the treatments, $\Psi \in \mathbb{R}^{G \times G}$ denotes a matrix that encodes covariances among the genes, and $\mu \in \mathbb{R}^{P \times G}$ denotes the mean.  The latent variable $\theta$ is then modelled using a Kronecker product:
\begin{equation}\label{eq:thetaprior}
\theta \sim \mathcal{N}(\mu, K \otimes \Psi).
\end{equation}
For $r \in \{1,2,\ldots, R\}$, let each $X^{(r)}$ given $\theta$ be an independent and identically distributed Gaussian:
\begin{align}\label{eq:datamodel}
X^{(r)}|\theta &\sim \mathcal{N}(\theta,\Lambda),
\end{align}
where $\Lambda \in \mathbb{R}^{PG \times PG}$ is a hyperparameter.

The Kronecker structure of the covariance of $\theta$ in \cref{eq:thetaprior} constrains the distribution of $\theta$, but without additional constraints the model defined by \cref{eq:thetaprior} and \cref{eq:datamodel} is still too flexible to be identified without a large number of replicates.  For example, treating $\Lambda$ as a free parameter, maximum likelihood estimation cannot be performed unless $R>PG$ because many hyperparameter values lead to infinite likelihoods.  However, in most widely used public cell-perturbation datasets, $R$ is at most three.  Therefore, it is necessary to constrain the hyperparameters such that maximum likelihood can be used to set the hyperparameters.  

We begin by reviewing some properties of Kronecker products and introduce the notation that will be necessary to define the hyperparameters (\cref{subsec:kroneckerfun}).  We then discuss ways to constrain $K$ (\cref{subsec:definingK}).  Using ideas from factor analysis, we next discuss ways to constrain $\Psi$ and $\Lambda$ (\cref{subsec:definingPsiLambda}).  Finally, we present an algorithm for finding the maximum-likelihood hyperparameters subject to these constraints (\cref{subsec:likopt}).

\subsection{Kronecker product notation}

\label{subsec:kroneckerfun}

\newcommand{\basisvec}[2]{{\varepsilon}_{#1,#2}}

We use Kronecker products to define our model, and use their algebraic properties to design computationally efficient algorithms.  To describe certain algebraic manipulations of Kronecker products, we here introduce some additional notation.  Let $\basisvec{N}{1},\ldots,\basisvec{N}{N}$ denote the standard basis in $\mathbb{R}^N$.  Let $\mathbf{1}_N = \sum_{i} \basisvec{N}{i}$ denote the all-ones vector and $I_{N} = \sum_i \basisvec{N}{i} \basisvec{N}{i}^\top$ denote the identity matrix.  In this notation, $(\basisvec{P}{p}^\top \otimes \basisvec{G}{g}^\top) \theta = \theta_{p,g}$.  


\subsection{Defining the treatment-similarity kernel \maybetexorpdf{$K$}{K} using embeddings}

\label{subsec:definingK}

We define $K$ with the squared exponential (SE) kernel:
\begin{align}\label{eq:Kdef}
K_{pq} = \sigma_p \sigma_q \exp(-\Vert \anembedding_p - \anembedding_q \Vert^2).
\end{align}
Here, $\sigma \in \mathbb{R}^P$ denotes the marginal variance and $\epsilon \in \mathbb{R}^{P \times H}$ denotes the positions used to construct the SE kernel.   In the literature on representation for small chemicals, each $\epsilon_p$ is known as the ``embedding'' for treatment $p$ \citep{sabando2022using}.  These embeddings can be rescaled versions of user-supplied embeddings $\anembedding'$.  Examples of such user-supplied embeddings are given in \cref{sec:realdata}.  We present two ways these user-supplied embeddings can be rescaled.  

The first way sets $\anembedding = \alpha \anembedding'$, where $\alpha\in \mathbb{R}$ defines a single lengthscale for the SE kernel.  The lengthscale controls $K$ as follows: as $\alpha$ grows smaller, $K$ indicates greater correlations between pairs of nearby perturbations.  If the distance between $\anembedding_p'$ and $\anembedding_q'$ is $\delta$, for example, then $K_{p,q}$ is $\sigma_p \sigma_q \exp(-\alpha \delta^2)$.   The hyperparameter $\alpha$ can be estimated via maximum likelihood, allowing us to learn which distances in the embedding space correspond to which covariances in $K$.  

The second way allows greater adaptivity by setting $\anembedding_{pi} = \alpha_i \anembedding_{pi'}$ where $\alpha \in \mathbb{R}^H$ gives per-coordinate lengthscales for the SE kernel.  This approach is known as automatic relevance determination \citep{rasmussen2005gaussian}.

\subsection{Setting gene-gene covariances \maybetexorpdf{$\Psi$}{} and measurement noise \maybetexorpdf{$\Lambda$}{}}

\label{subsec:definingPsiLambda}

We consider two ways to set $\Psi$ and $\Lambda$.  The first way sets $\Psi$ to be the identity and $\Lambda$ to be diagonal.  This is a computationally attractive option when $G$ is large, but it does not model dependence among the columns of $\theta$.  

The second way models additional covariance structure by positing that $\theta$ is low-rank.  We formulate this low-rank structure by introducing a latent variable $\tilde Z \in \mathbb{R}^{P \times L}$ and an orthogonal matrix hyperparameter $V \in \mathbb{R}^{G \times L}$.  The low-rank structure is ensured by setting $\theta = \tilde Z V^\top$.  For consistency with the literature on factor analysis, we refer to $\tilde Z$ as the ``treatment loadings'' and $V$ as the ``gene loadings.'' We model $\tilde Z$ using hyperparameters $\mu' \in \mathbb{R}^{PL}$ and $\psi \in \mathbb{R}^L$ as
\begin{equation} \label{eq:tildeZdef}
\tilde Z \sim \mathcal{N}(\mu', K \otimes \mathrm{diag}(\psi)).
\end{equation}
Under our specification that $\theta = \tilde Z V^\top$, the prior mean of $\theta$ is $\mu = (I_P \otimes V) \mu'$ and the gene-gene covariance hyperparameter is $\Psi = V \mathrm{diag}(\psi)V^\top$.  

In this second way, $\Lambda\in \mathbb{R}^{PG\times PG}$ hyperparameter is set using the same low-rank structure that defines $\Psi$. Specifically, we introduce additional latent per-replicate variables, $Z^{(r)} \in \mathbb{R}^{P \times L}$. Letting $\tau>0$ denote an isotropic noise level, we model the data as
\begin{equation}
X^{(r)}|Z^{(r)} \sim \mathcal{N}((I_P \otimes V) Z^{(r)},\tau^2 I_{PG}),
\end{equation}
where
\begin{equation} \label{eq:Zdef}
Z^{(r)}|Z \sim \mathcal{N}(\tilde Z,I_P\otimes \mathrm{diag}(\lambda)).
\end{equation}
We refer to $Z^{(r)}$ as the replicate-level treatment loadings.  The latent variables can be marginalized out to yield that
\begin{align}
\Lambda = (I_P \otimes V \mathrm{diag}(\lambda) V^\top) + \tau^2 (I_{P} \otimes I_G).
\end{align}

We next detail an algorithm for finding the maximum likelihood hyperparameters.

\subsection{Likelihood optimization with low-rank modelling}

\label{subsec:likopt}

To optimize the hyperparameters $\mu$, $K$, $\Psi$, and $\Lambda$ under the low-rank assumption described in \cref{subsec:definingPsiLambda}, we propose an expectation-maximization procedure.  There are two challenges to implementing expectation maximization in this context.

First, a naive approach to computing the expected log likelihood required by this procedure would require $O(P^3 L^3)$ operations per iteration.  This is prohibitively large for several of our datasets.  However, because $V$ is an orthogonal matrix, we can compute these expectations in $O(P^3 L)$ operations.  Note that the requirement that $V$ is orthogonal does restrict the expressivity of the model class; it implies that $\Psi$ and $V\mathrm{diag}(\lambda)V^\top$ can be diagonalized by the same eigenvectors $V$.  Indeed, for each $p$ and $r$, both $V^\top \theta_p$ and $V^\top (\theta_p - X^{(r)}_p)$ have diagonal covariances.  Thus, the orthogonality requirement enforces a common structure between our prior uncertainty about $\theta$ and the noise model $\Lambda$.  

Second, exact maximization of the expected log likelihoods cannot be performed in closed form.  Instead, we divide the parameters into three groups: $(\mu',K,\psi,\lambda)$, $V$, and $\tau^2$.  With any two of these groups fixed, the maximizer for the third group can be computed in closed form.  Therefore, we perform three block-coordinate ascent updates in each iteration of the procedure.  

\cref{alg:lowrank} outlines our expectation-maximization procedure.  The remainder of \cref{subsec:likopt} details each step of the algorithm.  

\begin{algorithm}
\begin{algorithmic}
\scriptsize
\Require{Data $X \in \mathbb{R}^{RPG}$}
\State $L\gets$ initial estimate based on $X$  \algorithmiccomment{Set the rank (\cref{subsubsec:initL})}
\State $\hat Z,V \gets$ singular vectors from $L$-truncated SVD \algorithmiccomment{Initialize loading estimates (\cref{subsubsec:initZhat})}
\State $\mu',K,\psi,\lambda \gets \argmax p_{\mu',K,\psi}(\hat Z)$ \algorithmiccomment{Initialize hyperparameters for treatment loadings (\cref{subsubsec:initmukpsi})}
\Repeat \algorithmiccomment{Expectation-maximization}
 \State $\hat \mu \gets \mathbb{E}_{\mu',K,\psi,\lambda,\tau,V}[Z | X]$, $\hat \Sigma \gets \mathrm{cov}_{\mu',K,\psi,\lambda,\tau,V}(Z | X)$  \algorithmiccomment{Calculate expectations (\cref{subsubsec:postexp})}
 \State $\mu',K,\psi,\lambda \gets \argmax \mathbb{E}_{\hat \mu, \hat \Sigma}[\log p_{\mu',K,\psi,\lambda,\tau,V}(Z,X)]$ \algorithmiccomment{Block-coordinate ascent (\cref{subsubsec:postmax})}
 \State $V \gets \argmin \sum_{rp} (X_{rpg} - \sum_k V_{gk} \hat \mu_{rpk})^2$ subject to $V^\top V=I_L$ \algorithmiccomment{Block-coordinate ascent (\cref{subsubsec:postmaxV})}
 \State $\tau^2 \gets (\Vert X - (I_{RP} \otimes V) \hat \mu \Vert^2 + \mathrm{tr}(\hat \Sigma)) / RPG$ \algorithmiccomment{Block-coordinate ascent (\cref{subsubsec:postmaxtau})}
\Until{convergence}
\end{algorithmic}
\caption{Estimating hyperparameters $L,\mu',K,\psi,\lambda,\tau,V$ under low-rank assumptions \label{alg:lowrank}}
\end{algorithm}

\subsubsection{Obtaining an initial estimate for the rank \maybetexorpdf{$L$}{L}\optionalperiod{}}   \label{subsubsec:initL} We first view $X \in \mathbb{R}^{RPG}$ as an $RP \times G$ through a reshaping operation.  We then construct a mask $S \in \{0,1\}^{RP \times G}$ by selecting $10\%$ of the entries uniformly at random to be one and all the others to be zero.  For each value of $L'\in \{1,\ldots,100\}$, we solve the optimization problem
\begin{align}
\min_{\hat Z_{L'},\hat V_{L'}} \sum_{(r,p,g):\ S_{r,p,g}=1} \left(X_{p,g}^{(r)} - \sum_{\ell=1}^{L'} \hat Z_{L',p\ell}^{(r)}\hat V_{L',g\ell}\right)^2.
\end{align}
For each $L'\in \{1,\ldots,100\}$ we then compute a corresponding loss on the masked entries,
\begin{equation}
\sum_{(r,p,g):\ S_{r,p,g}=0} \left(X_{p,g}^{(r)} - \sum_{\ell=1}^{L'} \hat Z_{L',p\ell}^{(r)}\hat V_{L',g\ell}\right)^2,
\end{equation}
and set $L$ to the value of $L'$ associated with the smallest held-out loss.  

\subsubsection{Initializing estimates for loadings \maybetexorpdf{$\hat Z$ and $V$}\optionalperiod{}}  \label{subsubsec:initZhat}  To form initial estimates of loadings $\hat Z$ and $V$, we view $X \in \mathbb{R}^{RPG}$ as an $RP \times G$ matrix and compute the $L$-truncated singular value decomposition $X \approx U \mathrm{diag}(e) V^\top$.  We set $\hat Z = U \mathrm{diag}(e)$.  

\subsubsection{Optimizing prior parameters using initial estimates\optionalperiod{}}  \label{subsubsec:initmukpsi} We fit hyperparameters $\mu'$, $K$, $\psi$, and $\lambda$, which govern the replicate-level treatment loadings $\hat Z$, by optimizing $p(\hat Z)$ using gradient descent.  To perform gradient descent, we must be able to compute $\log p(\hat Z)$ efficiently.  Fortunately, due to the structure of $\Psi$ and $\Lambda$, this quantity can be expressed as a sum of $L$ independent terms:
\begin{equation}
\log p_{\mu',K,\psi,\lambda}(\hat Z)=\sum_{k=1}^{L}\log\mathcal{N}\left(\hat{Z}_{\cdot\cdot k};\mu_{\cdot k}',\lambda_{k}(I_R \otimes I_P)+\psi_{k}\left(\mathbf{1}_{R}\mathbf{1}_{R}^{\top}\right)\otimes K\right).
\end{equation}
Here, we use the notation $\hat Z_{\cdot \cdot k} \in \mathbb{R}^{P \times L}$ to signify $(I_R \otimes I_P \otimes \basisvec{L}{k}^\top )Z$ and $\mu_{\cdot k}'$ to signify $(I_R \otimes \basisvec{L}{k}^\top) \mu'$.

\subsubsection{Computing posterior expectations\optionalperiod{}}  \label{subsubsec:postexp} The latent random variable $Z \in \mathbb{R}^{R \times P \times L}$ given $x$ follows a normal distribution that we will denote $\mathcal{N}\left(\hat{\mu}(x),\hat{\Sigma}\right)$.  The parameters of this normal distribution can be expressed using Kronecker products:
\begin{align}
\hat{\Sigma}^{-1} =
\frac{1}{\tau^{2}}(I_{RP}\otimes V^{\top}V) + 
\sum_{k=1}^{L}\left(\lambda_{k}I_{RP}+\phi_{k}\left(\mathbf{1}_{R}\mathbf{1}_{R}^{\top}\right)\otimes K\right)\otimes \basisvec{L}{k}\basisvec{L}{k}^\top
\end{align}
and
\begin{align}
\hat{\mu}(x)=&(\mathbf{1}_{RP}\otimes V)\mu+\frac{1}{\tau^{2}}\hat{\Sigma}\left(\left(I_{RP}\otimes V^{\top}\right)x-(\mathbf{1}_{RP}\otimes V)\mu'\right).
\end{align}
From the orthonormality requirement that $V^\top V=I_L$, it follows that
\begin{equation}
I_{RP}\otimes V^{\top}V=(I_{RP}\otimes I_L)= \sum_k I_{RP} \otimes \basisvec{L}{k}\basisvec{L}{k}^\top.
\end{equation}
Thus,
\begin{align}
\hat{\Sigma}^{-1} = 
\sum_{k=1}^{L}\left(\frac{1}{\tau^2}I_{RP}+\lambda_{k}I_{RP}+\phi_{k}\left(\mathbf{1}_{R}\mathbf{1}_{R}^{\top}\right)\otimes K\right)\otimes \basisvec{L}{k}\basisvec{L}{k}^\top.
\end{align}
We define $\Xi_k = \left(\lambda_{k}I_{RP}+\phi_{k}\left(\mathbf{1}_{R}\mathbf{1}_{R}^{\top}\right)\otimes K+\frac{1}{\tau^{2}}I_{RP}\right)$ so that $\hat{\Sigma}^{-1}=\sum_{k=1}^{L} \Xi_k \otimes \basisvec{L}{k}\basisvec{L}{k}^\top$.  To compute $\hat \Sigma \in \mathbb{R}^{RPL \times RPL}$, we must invert this sum.  Ordinarily, inverting an $RPL \times RPL$ matrix would require $O(R^3 P^3 L^3)$ operations.  However, $\hat \Sigma$ has a simple form: 
\begin{align} \label{eq:sigissoeasy}
\hat{\Sigma}=\sum_{k=1}^{L} (\Xi_k)^{-1} \otimes \basisvec{L}{k}\basisvec{L}{k}^\top.
\end{align}
This fact may be shown by multiplying $\hat \Sigma^{-1}$ by the right-hand side of \cref{eq:sigissoeasy}:
\begin{align}
\left(\sum_{k=1}^{L} \Xi_k \otimes \basisvec{L}{k}\basisvec{L}{k}^\top\right)\left(\sum_{j=1}^{L} (\Xi_j)^{-1} \otimes \basisvec{L}{j}\basisvec{L}{j}^\top\right) &= \sum_{j,k} \Xi_k (\Xi_j)^{-1} \otimes \basisvec{L}{k}\basisvec{L}{k}^\top\basisvec{L}{j}\basisvec{L}{j}^\top\\
&= \sum_{j,k} \Xi_k (\Xi_j)^{-1} \otimes \basisvec{L}{k}\basisvec{L}{j}^\top \mathbb{I}_{j=k} = I_{RPL}.
\end{align}
The constituent matrices $\Xi_1,\ldots \Xi_L$ can be computed in $O(R^3 P^3 L)$ operations.

\subsubsection{Updating prior hyperparameters for treatment loadings\optionalperiod{}}  \label{subsubsec:postmax}  Following the expectation-maximization algorithm, we must here optimize $\mathcal{L}(\mu',K,\psi)=\mathbb{E}_{\hat \mu, \hat \Sigma}[\log p_{\mu',K,\psi,\lambda,\tau,V}(Z,X)]$.   This objective has two parts:
\begin{align}
\mathcal{L}_{\mathrm{prior}} & =\mathbb{E}_{\hat \mu, \hat \Sigma}[\log p_{\mu',K,\psi}(Z)]\\
\end{align}
and
\begin{align}
\mathcal{L}_{\mathrm{data}} & =\mathbb{E}_{\hat \mu, \hat \Sigma}\left[\log\mathcal{N}\left(X;(I_{RP}\otimes V)Z,\tau^{2}I_{RPG}\right)\right].
\end{align}
We can ignore $\mathcal{L}_{\mathrm{data}}$ in fitting $\mu$, $K$, $\psi$, and $\lambda$, as $\mathcal{L}_{\mathrm{data}}$ is constant with respect to these parameters.  To compute $\mathcal{L}_{\mathrm{prior}}$, we exploit the independence structures in both $\Psi$ and $\hat \Sigma$ to find that 
\begin{equation}
\mathcal{L}_{\mathrm{prior}}  =\sum_{k}^L \mathbb{E}_{\hat \mu_{\cdot \cdot k}, \Xi_k^{-1}}\left[ 
 \log\mathcal{N}\left(Z_{\cdot\cdot k};\mu_{\cdot k}',\lambda_{k}I_{RP}+\psi_{k}\left(\mathbf{1}_{R}\mathbf{1}_{R}^{\top}\right)\otimes K\right)
\right].
\end{equation}
This summation can be computed in $O(LR^3P^3)$ operations. 

\subsubsection{Updating \maybetexorpdf{$V$}{V}\optionalperiod{}}  \label{subsubsec:postmaxV} We again seek to maximize $\mathbb{E}_{\hat \mu, \hat \Sigma}[\log p_{\mu',K,\psi,\lambda,\tau,V}(Z,X)]$, now with respect to $V$. 
 By dropping terms irrelevant to $V$, we find this optimization problem has the same optimizer as
\begin{equation}
\min \sum_{rp} (X_{rpg} - \sum_k V_{gk} \hat \mu_{rpk})^2\qquad \mathrm{s.t.}\ V^\top V=I_L.
\end{equation}
The latter problem is a Procrustes problem and therefore can be solved in $O(GL\max(G,L))$ operations by taking the product of the left and right singular vectors of the matrix $M$, defined by $M_{gk} = \sum_{rp} \hat \mu_{rpk} X_{rpg}$ \citep{gower2004procrustes}.  

\subsubsection{Updating measurement noise \maybetexorpdf{$\tau$}{}\optionalperiod{}}  \label{subsubsec:postmaxtau} We again maximize $\mathbb{E}_{\hat \mu, \hat \Sigma}[\log p_{\mu',K,\psi,\lambda,\tau,V}(Z,X)]$, this time with respect to $\tau$.  The optimizer is 
\begin{equation} \label{eq:tauupdate}
\tau^2 \gets \frac{\Vert X - (I_{RP} \otimes V) \hat \mu \Vert^2 + \mathrm{tr}(\hat \Sigma)}{RPG}.
\end{equation}

\ignoreme{

 ____    _  _____  _    ____  ____  _     ___ _____ 
|  _ \  / \|_   _|/ \  / ___||  _ \| |   |_ _|_   _|
| | | |/ _ \ | | / _ \ \___ \| |_) | |    | |  | |  
| |_| / ___ \| |/ ___ \ ___) |  __/| |___ | |  | |  
|____/_/   \_\_/_/   \_\____/|_|   |_____|___| |_|

}

\section{A data-splitting evaluation method}

\label{sec:datasplit}

Type S errors are particularly salient for cell-perturbation experiments.  Biologists must interpret the findings of such experiments in terms of gene control circuits.  Each circuit is understood in terms of upregulation and downregulation of genes \citep{davidson2001genomic}.  We propose a new approach for assessing type S errors of ATE estimators for cell-perturbation datasets.  This new approach has two key features: it does not require access to ground-truth ATEs, which are unavailable, and it does not require the estimators to be based on correctly specified models.    

The key idea behind our approach is to assess the \emph{proportion} of parameters for which an estimator makes a type S error, rather than trying to determine the probability of a type S error for each parameter individually.  This proportion summarizes the error of an estimator for many parameters using a single number.  In some cases, this single number may give an insufficiently detailed view of the errors made by an estimator.  To strike a balance between simplicity and detail, we therefore assess the type S error proportion for many different subsets of parameters. 

For any subset of treatment-gene pairs $\mathcal{S}$, let
$V_{\mathcal{S}}$ denote the proportion of parameters indexed by $\mathcal{S}$ for which the smoothed estimator $\theestimator$ made a type S error; i.e.,
\begin{equation}
V_{\mathcal{S}} = \frac{|\{(p,g) \in \mathcal{S}:\ \mathrm{sign}(\theta_{p,g}^*) \neq \mathrm{sign}(\theestimator_{p,g})\}|}{|\mathcal{S}|}.
\end{equation}
If the ground truth for $\theta^*_{p,g}$ were known for a representative subset $\mathcal{S}$ of treatments and genes, we could calculate this quantity directly.  However, ground truth is generally unavailable.

In this section, we propose a method both for assessing type S error without access to ground truth (\cref{assessing}) and for controlling it (\cref{controlling}). This method is applicable both to validating the estimator introduced in \cref{sec:estimator} and more generally.

\subsection{Assessing error}
\label{assessing}

To assess error, we require two ingredients.  First, we need an estimator with the following ``sign-validity'' property: the median of the estimator for $\theta^*_{p,g}$ must have the correct sign for each $p,g$.  We will denote this valid estimator by $\strawestimator$.  Second, we need at least two experimental replicates.  In typical cell-perturbation experiments, the number of replicates $R$ is two or three \citep{subramanian2017next,srivatsan2020massively,schmidt2022crispr}.  

Given these ingredients, we propose to assess error as follows.  First, split the replicates into two groups: $\{1,2,\ldots \tilde R\}$ and $\{\tilde R+1,\ldots, R\}$.  Then, compare the smoothed estimator based on the first group of replicates, $\theestimator(X^{(1)},\ldots X^{(\tilde R)})$, with the sign-valid estimator based on the second group of estimates, $\strawestimator(X^{(1)},\ldots X^{(\tilde R)})$.  Next, use $\strawestimator$ to bound the type S error rate of $\theestimator$ in terms of the Cross-replicate Sign Proportion (CSEP) for a subset $\mathcal S$ of the treatment-gene pairs:
\begin{equation}
\mathrm{CSEP}_{\mathcal{S}} = \frac{|\{p,g \in \mathcal{S}:\ \mathrm{sign}(\strawestimator_{p,g}) \neq \mathrm{sign}(\theestimator_{p,g})\}|}{|\mathcal{S}|}.
\end{equation}
In words, the CSEP for $\mathcal S$ is the proportion of parameters indexed by $\mathcal S$ for which the sign of the estimator based on the first group of replicates disagrees with the sign of the estimator based on the second group of replicates. 

For any subset of treatment-gene pairs $\mathcal{S}$, \citet{loper2022quantifying} show that the error proportion
\begin{equation}\label{eq:vcsepbound}
V_\mathcal{S} \leq 2 \cdot \mathbb{E}[\mathrm{CSEP}_\mathcal{S}].
\end{equation}
This upper bound holds regardless of the form of $\theestimator$, giving us a model-free method for evaluating its type S error rate.

\subsection{Controlling error}
\label{controlling}

\cref{eq:vcsepbound} can also be used to control type S error, i.e., to estimate a subset of parameters $\mathcal S$ in which the smoothed estimator has the targeted type S error proportion.  To do so, first construct a nested family of subsets of treatment-gene pairs, $\mathcal{S}_1\subseteq \cdots \subseteq{S}_{K}$.
Next, compute the CSEP for each subset.  Finally, select the largest subset where the corresponding CSEP lies below $2V^*$.

In the case studies that follow, we take $\mathcal{S}_{k} = \{(p,g):\ \theestimator_{p,g} \leq \theestimator_{(k)}\}$ where
$\theestimator_{(1)},\ldots,\theestimator_{(PG)}$ are
order statistics of the magnitudes of the smoothed estimates.
We plot $\mathrm{CSEP}_{\mathcal{S}_k}$ against $|\mathcal{S}_k|$ to visualize error bounds of various estimators.  

\ignoreme{
 ____            _   ____        _            
|  _ \ ___  __ _| | |  _ \  __ _| |_ __ _ ____
| |_) / _ \/ _` | | | | | |/ _` | __/ _` |_  /
|  _ <  __/ (_| | | | |_| | (_| | || (_| |/ / 
|_| \_\___|\__,_|_| |____/ \__,_|\__\__,_/___|
                                              
                                              }

\section{Case studies}

\label{sec:realdata}

We apply our smoothed estimator (\cref{sec:estimator}) to data from cell-perturbation experiments and use the CSEP (\cref{sec:datasplit}) to evaluate whether the smoothed estimator improves accuracy.  The first case study shows how eight different experimental conditions can be represented through embeddings that are suitable for the kernel from \cref{eq:Kdef}.  We evaluate the performance of the corresponding smoothed estimator (\cref{subsec:crisprdata}).  In the second case study, each treatment is associated with a different small chemical.  We devise our smoothed estimators by embedding these chemicals into a Euclidean space using Mordred fingerprints and using the low-rank kernels and optimization procedures developed in \cref{subsec:likopt}.  In both case studies, we find that the smoothed estimator outperforms the alternatives.   

\subsection{Smoothing across experimental conditions}

\label{subsec:crisprdata}

To infer gene regulatory networks, \citet{schmidt2022crispr} used CRISPRa to increase the expression of various genes and observe how the expression of other genes changed in response.

The procedure of \citet{schmidt2022crispr} estimates a large number of parameters.  These parameters can be organized into a fifth-order tensor with shape $2\times 2\times 2 \times 9 \times 24,709$.  The first three modes of this tensor correspond to binary choices about the experimental design, the fourth corresponds to the gene targeted for intervention, and the fifth corresponds to the gene whose expression is measured.  The first choice (CD4 vs. CD8) relates to the population under study.  Before donor cells are perturbed, they are either filtered so that they mostly include a type of T cell known as CD4 or they are filtered to mostly include a type of T cell known as CD8.  The second choice (guide 0 vs. guide 1) reflects that the CRISPR augmentation technology increases a gene's expression by targeting a particular locus in the gene.  \citet{schmidt2022crispr} investigated two different loci for each target gene.  The third choice (unstimulated vs. stimulated) reflects whether an additional stimulation step was performed prior to measurement.   This stimulation step activates certain functions of T cells by introducing additional small chemicals into the cultures.   In each of two replicates, \citet{schmidt2022crispr} constructs a pool of cells with each cell type, guide choice, stimulation state, and gene target. The expression values for $24,709$ different genes are then measured for each cell in the pool.  For each replicate, cell type and stimulation state \citet{schmidt2022crispr} also constructs a pool of control cells and measures gene expressions for those cells. 

We estimate the ATEs using the smoothed estimator from \cref{sec:estimator}.  To use this estimator, we first reshape the parameters of interest into a matrix of shape $8 \times 222,381$.  The first mode of the original tensor (with shape $2 \times 2 \times 2$) forms the rows and the last two modes (with shape $9 \times 24,709$) form the columns.  We construct an embedding for each row using one-hot vectors.  For example, $\anembedding_p'=(0,0,1)$ indicates that treatment $p$ used guide 0 on stimulated CD4 cells.  As described in \cref{subsec:definingK}, we use a $3$-dimensional automatic relevance determination kernel to specify $K \in \mathbb{R}^{8 \times 8}$.  We learn $\sigma \in \mathbb{R}^8$ as a free parameter, set $\Phi=I_{24,709 \cdot 9}$, and set $\Lambda = \mathrm{diag}(\lambda) \otimes I_{24,709 \cdot 9}$ where $\lambda \in \mathbb{R}^8$ is a tuneable parameter.    In both replicates, we obtain a measurement $X^{(r)}_{p,g}$ by calculating a $z$-score based on the Mann-Whitney statistic for the null hypothesis that the expression of gene $g$ in the cells perturbed by treatment $p$ is the same as the expression of gene $g$ in the corresponding control cells.  Such $z$-scores are asymptotically normal \citep{lehmann1951consistency}.  We use these $z$-scores in three ways: as the observed values used as a basis for our smoothed estimator, as a baseline estimator to be compared with the new smoothed estimator, and as sign-valid estimators.  


\begin{figure}
\begin{center}
\includegraphics[width=\figwidth]{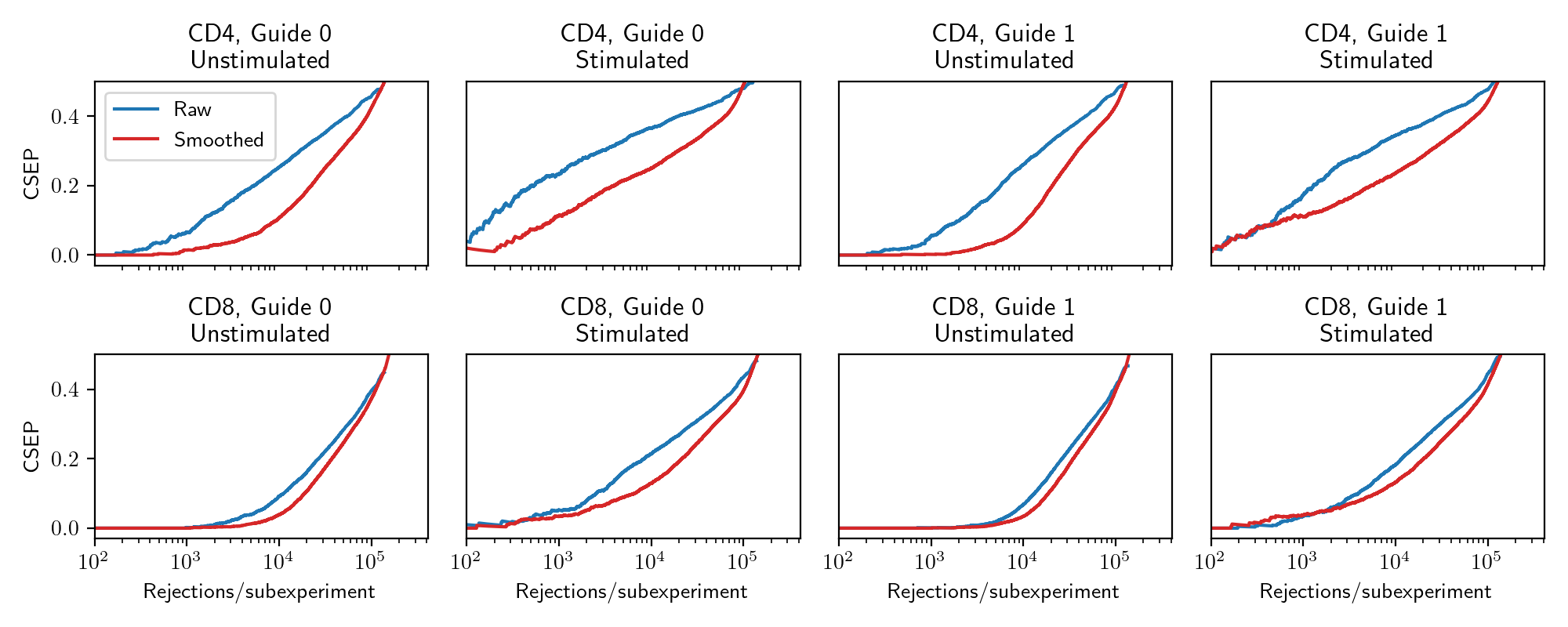}
\caption{
\captionheading{Our smoothed estimator improves accuracy in a CRISPR augmentation dataset.}  In real data, we cannot directly calculate estimator accuracy.  However, the Cross-replicate Sign Error Proportion (CSER) measures disagreement between an estimator and held-out data, and the type S error proportions are bounded by twice its expected value.
\label{fig:schmidt}}
\end{center}
\end{figure}

Figure \ref{fig:schmidt} shows that our smoothed estimator yields better type S error control than the raw estimator for every treatment.  The automatic relevance detection model in the smoothed estimator also yields insight into the consequences of various experimental choices.  The learned hyperparameters of this model include a coefficient for each choice (e.g., guide 0 versus guide 1).  A lower coefficient creates higher correlations (e.g., the ATEs of guide 0 are more similar to ATEs of guide 1).  We found coefficents of $0.25$ for the CD8 vs. CD4 choice, $0.31$ for the guide 0 vs. guide 1 choice, and $0.63$ for the stimulated vs. unstimulated choice.  


\subsection{Smoothing across the space of small chemicals}
\label{subsec:smallchem}

\citet{subramanian2017next} and \citet{srivatsan2020massively} both performed high-throughput cell-perturbation experiments with small chemicals, investigating how each small chemical affects gene expressions.  These small chemicals can be represented using Simplified Molecular-Input Line-Entry System (SMILES) strings \citep{o2012towards}, and these strings can be embedded as a matrix $\anembedding' \in \mathbb{R}^{P \times H}$ using Mordred fingerprints \citep{moriwaki2018mordred}.  

Using these embeddings, we apply our smoothed estimator to data from the L1000 and Sci-Plex protocols.  The L1000 protocol yields a dataset of $z$-scores.  The Sci-Plex protocol yields gene expressions for treated and control cells, and we obtain $z$-scores for this dataset using Mann-Whitney U statistics.  We use these $z$-scores as the noisy measurements ($X$) posited by the Bayesian model that defines our smoothed estimator.   For both protocols, we consider three estimators.  The first estimator, ``Raw,'' is computed by averaging the $z$-scores over replicates.  The second estimator, ``PCA,'' is computed by terminating \cref{alg:lowrank} early and returning $\sum_r \hat Z^{(r)}V^\top/R$.  This estimator uses the PCA-based initialization strategy but does not use Gaussian processes.  The third estimator, ``Smoothed,'' is computed using all of \cref{alg:lowrank}.  In Sci-Plex data, the number of replicates $R$ is two, and we split the data by taking one replicate for testing and one for training.  In the L1000 data, there are three replicates; we use two replicates for training and one for testing.

\begin{figure}
\begin{center}
\includegraphics[width=\figwidth]{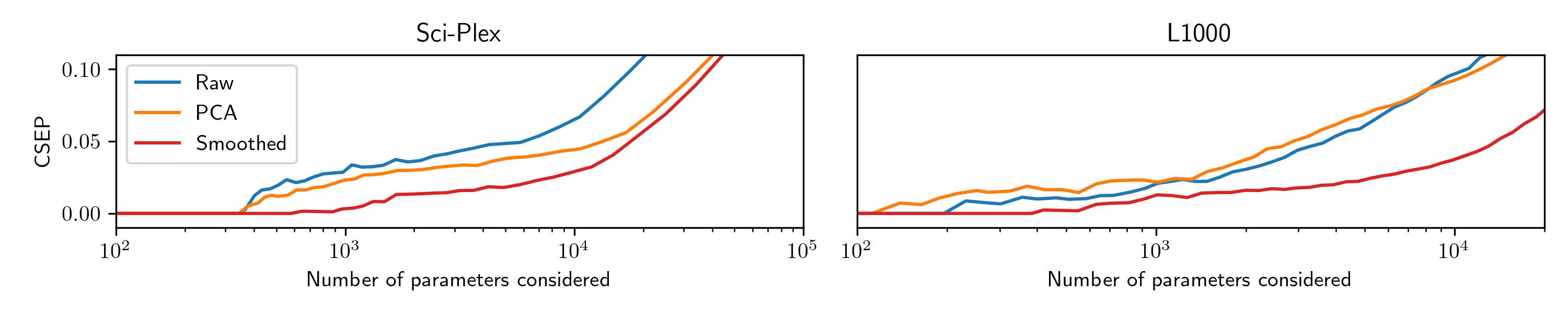}
\caption{
\captionheading{Our smoothed estimator has higher accuracy on small chemical perturbation datasets.}  As in \cref{fig:schmidt}, Cross-replicate Sign Error Proportions (CREPs) are used to calculate upper bounds on error rates. 
\label{fig:sciplex}}
\end{center}
\end{figure}

Figure \ref{fig:sciplex} compares all three estimators using the data-splitting assessment, formed using the raw $z$-scores as sign-valid estimators.   The smoothed estimator always attains the highest CSEP.  For example, targeting a CSEP of 5\% (corresponding to a type S error proportion of 10\%), the smoothed estimator yields 2.5 times more discoveries on Sci-Plex data and 3.5 times more discoveries on L1000 data.

\ignoreme{
 ____ ___ __  __ ____  
/ ___|_ _|  \/  / ___| 
\___ \| || |\/| \___ \ 
 ___) | || |  | |___) |
|____/___|_|  |_|____/ 
                       
}

\section{Simulations and parameter identifiability}

\label{sec:sims}

In both case studies above, we set $\theta$ to follow a parametric prior and set its hyperparameters to maximize the likelihood of the data.  However, the data only includes a small number of independent replicates.    To investigate whether data scarcity could lead to poor performance, we developed semi-synthetic datasets based on the L1000 data from Section \ref{subsec:smallchem}.  In these semi-synthetic datasets, unlike in our case studies, we have access to the ground-truth ATEs.  This allows us to directly evaluate type S error proportions, whereas in real data we can only construct upper bounds on the type S error proportions.  

In all simulations, we assume $\theta \in \mathbb{R}^{P \times G}$ has a rank-10 structure and $\theta = \tilde Z^* (V^*)^\top$.  We choose $V^*$ and $Z^*$ to approximate realistic data, defining them in terms of the top 10 principal components of a subset of the L1000 dataset with $P=200$ perturbations and $G=978$ genes.  In our first two simulated datasets, we draw each observation $X^{(r)}$ by adding independent standard Gaussian noise to each entry of $\theta$; in the first dataset we set $R=1$ and in the second dataset we set $R=2$. In the third simulation we set $R=2$ and also introduce per-replicate batch effects.  Batch effects are thought to have a low-rank structure \citep{zhang2022structured}.  Therefore, for each replicate $r$, we draw matrices $\check Z^{(r)} \in \mathbb{R}^{P\times 10}$ and $\check V^{(r)} \in \mathbb{R}^{P \times 10}$ using i.i.d.~draws from normal distributions.  We then use these matrices to produce our third simulation dataset by sampling
\begin{align}
X^{(r)} \sim \mathcal{N}\left(\theta + \frac{1}{2\sqrt{10}}\check Z^{(r)} (\check V^{(r)})^\top,\left(\frac{3}{2}\right)^2 I_{P\times G}\right).
\end{align}

For each of the three simulations we consider three estimators of $\theta$: ``Raw,'' ``PCA,'' and ``Smoothed.''  These estimators are as described in \cref{subsec:smallchem}.  We explore two choices for embeddings in our smoothed estimators.   The ``Smoothed (uninformative distance)'' estimator embeddings come from $P$ points in $\mathbb{R}^{10}$ drawn from standard normal distributions.  The ``Smoothed (informative distance)'' embeddings are given by the rows of $\tilde Z^*$.

For each simulation and each method, we consider two metrics of estimator performance.  The first metric assesses the type S error proportion within different subsets of parameters.  In particular, for every threshold $t$, we consider the set $\mathcal{S}$ of parameters $(p,g)$ where $|\hat \theta_{p,g}|>t$ and compute the type S error proportion in that subset, $V_\mathcal{S}$.   The results of these computations can be visualized as a graph plotting the number of parameters in the set associated with threshold $t$ against the corresponding type S error proportion.  The second metric we consider is the correlation coefficient between the vector $(\hat \theta_{p,1},\ldots\hat \theta_{p,G})$ and the vector $(\theta_{p,1},\ldots\theta_{p,G})$ for each estimator $\hat \theta$ and each perturbation $p \in \{1\ldots P\}$.  These coefficients can be visualized using box plots.  

\begin{figure}
\begin{center}
\includegraphics[width=\figwidth]{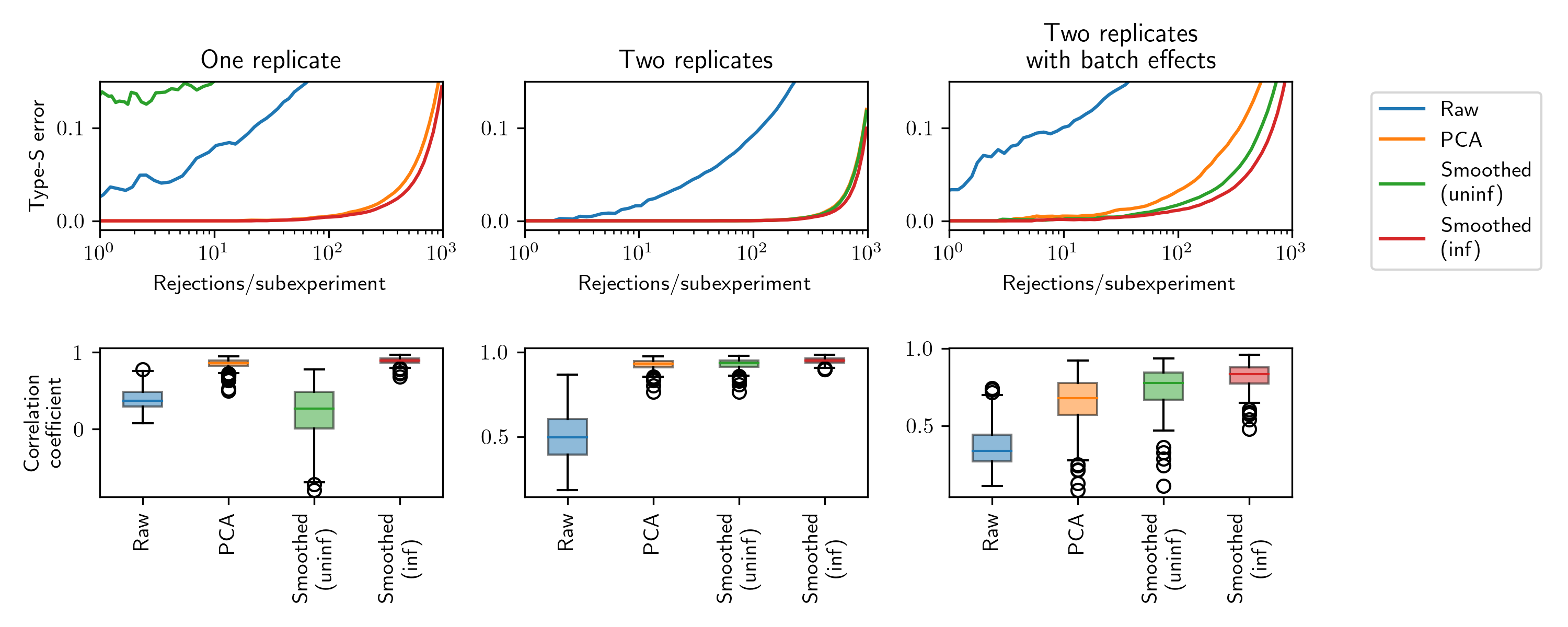}
\caption{
\captionheading{Comparing four estimators: raw $z$-scores (``Raw''), PCA approximated $z$-scores (``PCA''), smoothed estimators based on $z$-scores and uninformative embeddings (``Smoothed (uninf)''), and smoothed estimators based on $z$-scores and informative embeddings (``Smoothed (inf)'').}
Given two experimental replicates, smoothing consistently reduces type S error and improves correlations.  However, if the user supplies uninformative embeddings and performs the experiment only once, smoothing can lead to poor results.
\label{fig:sims}
}
\end{center}
\end{figure}

The results are presented in Figure \ref{fig:sims}.  We have three main findings.  First, even when the embeddings are uninformative, the smoothed estimator is still the most accurate as long as at least two replicates are available.  Second, our smoothed estimator leads to even greater improvements if batch effects are present.  Finally, with only one replicate, our smoothed estimator can have poor performance if uninformative embeddings are provided.   

We conjecture that the poor performance of the smoothed estimator in the third setting is due to a non-identifiability that arises in the absence of replication.  Specifically, if $\ell=0$ the marginal covariance of each $Z^{(r)}$ becomes $I_P\otimes (\Psi + \Lambda)$, where $\Psi$ and $\Lambda$ are both free parameters.  Two distinct hyperparameter choices, $(\Psi,\Lambda)$ and $(\Psi',\Lambda')$, cannot be distinguished using data as long as $\Psi+\Lambda=\Psi'+\Lambda'$ and $R=1$.  Fortunately, we can often determine whether this issue applies to a given dataset, even if only one replicate is performed.  For example, we could conduct a hypothesis test with the null hypothesis that $\ell=0$.

\section{Discussion: towards multi-platform analysis}
\label{discussion}

The smoothed estimator developed in this paper is designed to analyze data from a single experimental platform (e.g., L1000 \emph{or} Sci-Plex).  However, because new platforms are continually being invented, the total amount of data available for any single platform is often much less than the data available across many platforms. Each individual platform is subject to different technical artifacts and limitations.  When several different platforms estimate overlapping sets of estimands, it may be possible to integrate measurements across all platforms to yield estimates with fewer type S errors.  Cross-platform integration has already shown promise in revealing key gene circuits from observational (i.e., non-interventional) data \citep{foltz2023cross}.  In the context of cell-perturbation experiments, integrating multiple platform's estimates of ATEs of the same treatments on the same genes may prove even easier than the task considered in this paper, namely, integrating estimates of ATEs for different treatments on the same set of genes.  The smoothed estimator proposed by this paper may be an effective tool in this context.  However, there are three difficulties that must be overcome to extend the methods in this paper to multi-platform setting: estimand mismatch, computational burden, and error mismatch.  

First, any multi-platform smoothing estimators must account for the fact that different platforms may have been designed to estimate subtly different estimands.  For example, L1000 measures bulk RNA expression among a large number of cells, whereas Sci-Plex measures RNA expression on a cell-by-cell basis.  As such, the estimands for L1000 correspond to changes in total expression over a population (in which cells with larger \emph{total} RNA counts will be disproportionately represented) whereas the estimands in Sci-Plex correspond to changes in average per-cell expression (in which all cells are represented equally).  The squared exponential kernels used in this paper may not adequately model such subtle distinctions.

Second, multi-platform smoothing estimators will require new computational methods.  The estimators in this paper depend on Gaussian processes, and the cubic computational scaling for inference with GPs can make it challenging to scale these approaches to larger datasets.  There are a variety of popular tools available for computational scaling, but further work is needed to adapt them to cell-perturbation data.  For example, GPyTorch \citep{gardner2018gpytorch} relies on accelerated hardware, such as GPUs with limited memory; it fails if this memory is exhausted.  KeOps \citep{charlier2021kernel} can mitigate these limitations by batching some computations and swapping information between accelerator and main memory, but it cannot readily exploit the Kronecker structure of our covariance kernels.  On the other hand, low-memory approaches such as KISS-GP \citep{wilson2015kernel} are inapplicable in our examples with high-dimensional embeddings.  KISS-GP requires that the kernel can be expressed as a Kronecker product of $H$ terms where $H$ is the dimension of the embedding.  The embedding space for Mordred fingerprints has millions of dimensions, which makes this approach infeasible.  To enable smoothing estimators that can incorporate more measurements, we must develop GP inference methods that are more suited for cell-perturbation data.

Third, although the data-splitting evaluation method proposed in this paper could be applied unchanged to multi-platform smoothing estimators, its utility would be limited because it focuses on unweighted error rates.  The type S error proportion bounded in this paper is the number of errors divided by the total number of estimands.  In multi-platform estimation, different platforms may have different numbers of estimands, and so this error metric would be dominated by platforms which estimate large numbers of parameters.  In this context, it may be more suitable to only consider errors for a subset of key genes that are measured across all platforms.  Measurements from other genes would serve to improve accuracy for estimates of the key genes, but would not be considered for the purposes of measuring error.  Such metrics may lead to alternative smoothed estimators that are more suitable for this regime.

\section{Conclusion}
\label{conclusion}

Cell-perturbation experiments offer great insight into gene regulatory networks.  However, these experiments are expensive, and typical estimators based on experimental data make many sign errors.  This paper makes two contributions to facilitate the analysis of this data.  First, we develop new ``smoothed'' estimators that use auxiliary information to reduce error.  Second, we develop a new method for assessing type S error without access to ground-truth ATEs.  There is a pressing need for such assessments in light of recent results that many model-based estimators give overinflated confidence \citep{li2022exaggerated}.  In real data from L1000, Sci-Plex, and CRISPRa-based platforms, the type S error assessments show that the smoothed estimators yield superior bounds on the error rates.  Smoothed estimators---when validated by a data-splitting error control procedure---offer a promising way to obtain more insight from these expensive experiments.
